\documentclass{article}
\usepackage{ijcai19}

\usepackage{times}
\usepackage{soul}
\usepackage{url}
\usepackage[hidelinks]{hyperref}
\usepackage[utf8]{inputenc}
\usepackage[small]{caption}
\usepackage{graphicx}
\usepackage{amsmath}
\usepackage{booktabs}
\usepackage{algorithm}
\usepackage{algorithmic}
\usepackage{verbatimbox}
\usepackage{float}
\floatname{algorithm}{Listing}
\title{Scalable Deployment of AI Time-series Models for IoT}

\author{
Bradley Eck\and
Francesco Fusco\and
Robert Gormally\and
Mark Purcell\and
Seshu Tirupathi
\affiliations
IBM Research Ireland
\emails
\{francfus, bradley.eck,seshutir,robertgo,markpurcell\}@ie.ibm.com
}
\begin{document}

\maketitle

\begin{abstract}
IBM Research Castor, a cloud-native system for managing and deploying large numbers of AI time-series models in IoT applications, is described. Modelling code templates, in Python and R, following a typical machine-learning workflow are supported. A knowledge-based approach to managing model and time-series data allows the use of general semantic concepts for expressing feature engineering tasks. Model templates can be programmatically deployed against specific instances of semantic concepts, thus supporting model reuse and automated replication as the IoT application grows. Deployed models are automatically executed in parallel leveraging a serverless cloud computing framework. The complete history of trained model versions and rolling-horizon predictions is persisted, thus enabling full model lineage and traceability. Results from deployments in real-world smart-grid live forecasting applications are reported. Scalability of executing up to tens of thousands of AI modelling tasks is also evaluated.
\end{abstract}

\section{Introduction}

Internet-of-Things (IoT) technology creates vast potential for improved decision making in several industries (e.g. smart grids, transportation systems, buildings management), based on access to large and timely information coming from increasing volumes of internet-connected sensing devices. Advances in machine learning and artificial intelligence (AI) provide for the necessary tools to make sense and learn from large amounts of time-series data using classical time-series models or more recent deep-learning approaches \cite{boxjen76,Rangapuram2018}. It is, however, widely recognised that the deployment of large quantities (thousands or even millions) of AI models in operational IoT systems is still a challenge, involving many manual, time-consuming tasks such as data exploration and preparation, or even handling the scalability of models execution and the persistence of model predictions. 

The use of a knowledge-based representation of IoT data has recently been investigated to enable automation of machine-learning tasks such as data exploration and feature engineering by leveraging semantic reasoning \cite{Chen2018,Zhang2017,Ploennigs2017}. In addition, the maturity of cloud computing has provided developers with the necessary technology to abstract away most operational concerns regarding the execution of code on the cloud. The serverless computing framework, in particular, has emerged as a new compelling paradigm for the deployment of cloud services \cite{Baldini2017}, with applications to time-series analysis \cite{Freeman2016} and image classification with deep-learning models \cite{Ishakian2018}.

Leveraging the key technologies of serverless computing and knowledge-based data representation, a system for the management and deployment of large quantities of AI time-series forecasting models in IoT applications was designed. The system, named IBM Research Castor, supports the deployment of custom AI modelling code (currently both in Python and R) on the cloud for automatic training and scoring according to user-defined schedules. 
A knowledge-based time-series data store allows the use of semantic reasoning in the model implementation, thus enabling the programmatic deployment of any given AI model to many specific instances of general semantic concepts (for example a model prepared for the prediction of electrical energy demand can then be deployed automatically to all specific  substations, buildings or other entities that consume electricity). 
The execution of the AI models leverages the serverless computing framework, thus providing for built-in parallelization and horizontal scalability features.  

The remainder of the paper is organised as follows. Section \ref{sec:overview} overviews the system architecture and workflow. The focus is then put on the AI model preparation and deployment, in Section \ref{sec:modelling}. Results from smart-grid deployments of the system, supporting live updates of hundreds of localised forecasts of energy demand and generation, are reported in Section \ref{sec:results}. Conclusive remarks are given in Section \ref{sec:conclusion}.

\section{IBM Research Castor Overview}\label{sec:overview}

\begin{figure*}[ht]
    \centering
    \includegraphics[width=15cm]{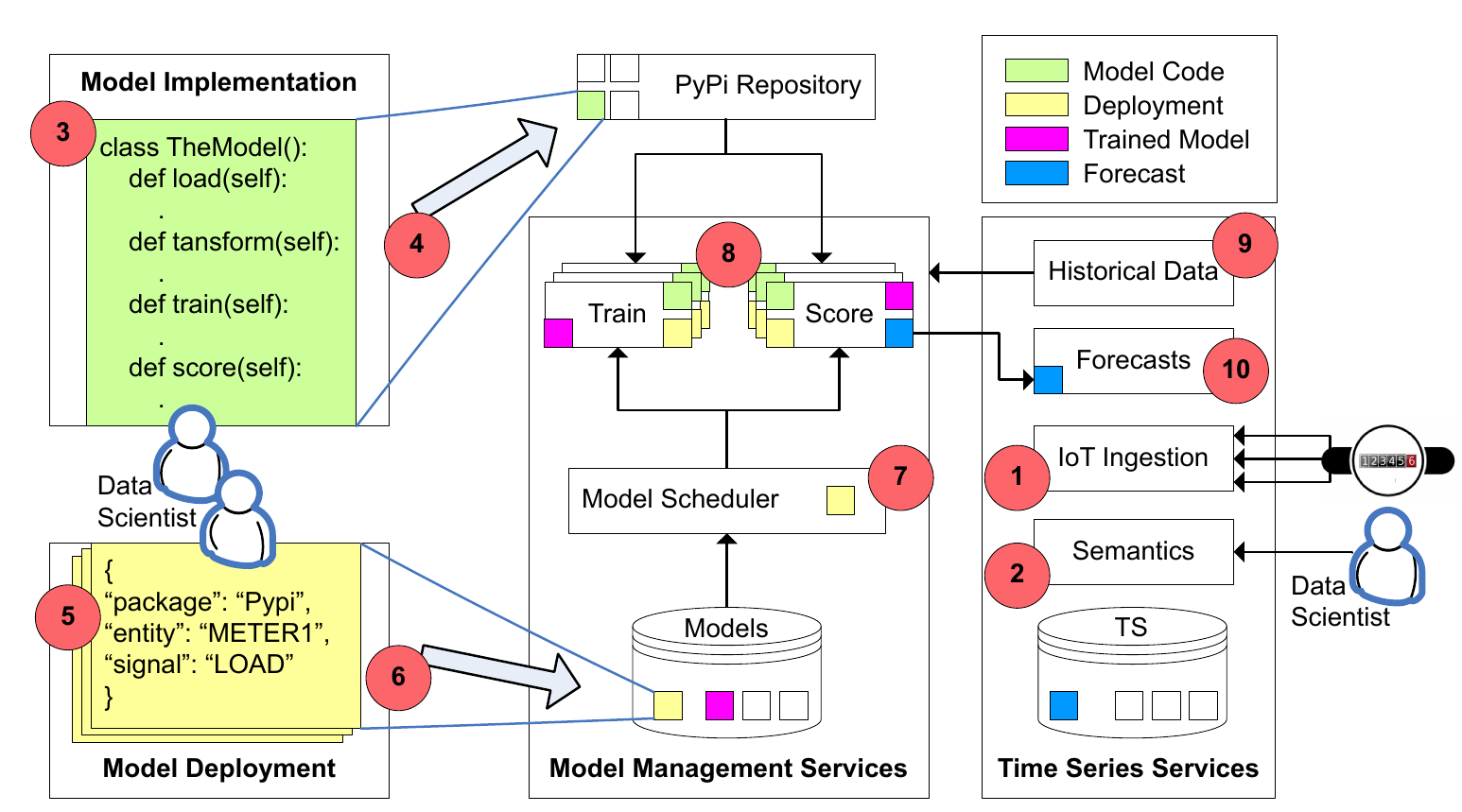}
    \caption{IBM Research Castor Overview and Modelling Flow}
    \label{fig:system_overview}
\end{figure*}

IBM Research Castor follows a micro-services architectural pattern with two main micro-service groupings: a time-series management and a model management suite of micro-services (see Figure \ref{fig:system_overview}). Both sets of micro-services and the supporting databases are deployed on the cloud and utilise serverless computing. Serverless is particularly suited to short-lived, bursty computational load that is typical of AI model training and scoring routines in IoT applications. Besides providing a more effective cloud resource allocation, only when a model is executed, the serverless framework removes the burden of server-side resource management and offers built-in parallelization and potentially infinite horizontal scalability. 

Invocations of certain micro-services are triggered based on real-world events, such as the availability of time-series data (1). The IoT data is ingested often at irregular frequencies, with many devices submitting data in parallel. Ingested time-series are stored in a knowledge-based data store. As new time-series become available, data scientists can provide additional semantics (2) to contextualise the time-series \cite{Chen2018}. IoT data ingestion and semantic definition occur out-of-band, and do not directly affect the main modelling work flow.

When individual time-series are understood by a data scientist, AI modelling software can be implemented to perform the tasks required by the IoT application, such as electricity demand time-series forecasting (3). A model implementation needs to load the data of interest, transform it into a usable in-memory representation, train a model based on historical data and produce a prediction from the trained model. Semantic reasoning can be used for expressing feature engineering and modelling steps in terms of concepts defining the application semantics. Upon completion, the model code (Python is assumed), is packaged and deployed to a PyPi repository (4). 

A model implementation (\emph{green}) can then be used for execution in many model deployments. Based on semantics (2) and desired model training/scoring schedules, model deployments (see Section \ref{sec:model_deployment}) are written (5) and registered with the system (6). The model deployments (\emph{yellow}) are stored in a database and specify the execution of the model implementations against specific instances of the application semantic concepts. A model scheduling micro-service (7) periodically loads the registered model deployments and determines if each model in question is due for training or scoring, based on the user-specified schedules. Model execution involves automatically installing the model implementation (\emph{green}) from the PyPi repository (8) and running the relevant model implementation routines (3) with specific model deployment configuration (5) and semantic data (2) loaded from the system data store. At model training, historical data (9) are retrieved from the time-series micro-service suite and a new \emph{model version} object (\emph{pink}) is produced, including a model object with fitted parameters (for example the weights of a neural network) and some training metadata, such as train time. The model version is saved into the data store and mapped to the semantic context of the model. At model scoring, the model version object is loaded from the system data store and made available to the scoring routine, which produces a time-series forecast (\emph{blue}). The forecast micro-service (10) handles the persistence of prediction time-series in a database.

\section{AI Time-series Model Management} \label{sec:modelling}

Creating an AI model for use on IBM Reserach Castor involves two separate steps of implementing the functionality itself and deploying the model by specifying the parameters that control its execution. In the following, Section \ref{sec:model_preparation} details the model implementation, based on a typical machine-learning workflow of load, transform, train and score. 
Section \ref{sec:model_deployment} discusses model deployment, where a model implementation is associated with a specific semantic context and  other configuration governing when and how the model should be executed.  
Separation between model implementation and deployment configuration is a key aspect of the system, designed to enable reuse and programmatic deployment of AI models.

\subsection{Model Implementation} \label{sec:model_preparation}

\begin{algorithm}[tb]
\caption{Model Implementation Pseudocode}
\label{alg:model_implementation_example}
\begin{verbnobox}[\small]
class MyModel (ModelInterface):
  def __init__ (self, context, task, 
                modelId, modelVersion, 
                user_params) 
  def load():
    x = getTimeseries(context.entity, 
                      context.signal, 
                      start, end) 
    w = getWeather(context.entity.lat, 
                   context.entity.long, 
                   start, end)
  def transform():
    x = merge(x, w)
    x = align_data (x, user_params.frequency)
    generate_lagged_features(x)
    // other features
  def train():
    start, end = user_params.train_period
    self.load(), self.transform()
    return( LinearRegression.fit(x)) 
  def score():
    start, end = now(),now()+dt(hours=24)
    self.load(), self.transform()
    model = getModel(modelID, modelVersion) 
    return( model.predict(x) ) 
\end{verbnobox}
\end{algorithm}

Model functionality is implemented as a collection of four functions:  \emph{load} data from available sources; \emph{transform} the data to prepare model features; \emph{train} the model according to the chosen algorithm; and \emph{score} a trained model to make predictions. The system imposes very few restrictions on these functions aside from the requirement that they work together.  Thus, load can retrieve data from our system or any other system accessible at run time. In practice these functions are grouped into a Python or R object of a special
class (Listing \ref{alg:model_implementation_example}).  The developer of a model places the program code for the class in a code repository. At run time the system retrieves the model and provides the execution parameters. 

As shown in Listing \ref{alg:model_implementation_example}, the model code has access to a number of parameters which become available at execution time: \emph{context} provides semantic information associated with the time-series targeted by a specific model instance, in terms of concepts of signal (what physical quantity, what unit) and entity (what location, what name, type, GIS coordinates,..); \emph{task} indicates whether the model should train or score; \emph{modelId} and \emph{modelVersion} point to the model and model version data (e.g. the weights of a trained neural network); \emph{user\_params} contains built-in parameters controlling model execution such as prediction window and time-step, as well as additional parameters that the user chooses to further customize the model implementation. The model context and any custom entries in \emph{user\_params} are fixed by the user at deployment time, as described in Section \ref{sec:model_deployment}, while the other parameters are transparently populated by the model execution engine, described in Section \ref{sec:overview}, and are not visible to the user. 

\subsubsection{Data Transformation Models}
The model implementation is quite flexible and can be used to carry out any other desired data processing tasks besides machine-learning time-series prediction. For example, in several deployments of our system, as shown in Section \ref{sec:results_data}, models perform data transformations such as creating regular energy time-series from integration and resampling of an irregular, instantaneous current or power data feed.


\subsection{Programmatic Deployment of AI Models} \label{sec:model_deployment}

\begin{algorithm}[tb]
\caption{Example of Model Deployment Configuration}
\label{alg:model_deployment_example}
\begin{algorithmic} 
\STATE \{
\STATE \; "context": \{"entity": \textless value\textgreater, "signal":\textless value\textgreater \},
\STATE \; "model\_name": \textless value\textgreater, 
\STATE \; "dist\_name": \textless myModelCodePackage\textgreater,
\STATE \; "dist\_ver": "1.0.0",
\STATE \; "module": \textless myModelCodeRoutines\textgreater,
\STATE \; "training\_deployment": \{
\STATE \qquad "time": "2019-03-01T00:00:00+00:00",
\STATE \qquad "repeatEvery": "1\_week" \},
\STATE \; "scoring\_deployment": \{
\STATE \qquad "time": "2019-03-01T00:00:00+00:00",
\STATE \qquad "repeatEvery": "1\_hours" \},
\STATE \; "user\_parameters": \{
\STATE \qquad "frequency": "15T,
\STATE \qquad "train\_time": \{"2018-01-01", "2019-01-01"\} \},
\STATE \}
\end{algorithmic}
\end{algorithm}

The \emph{deployment} of a model involves the creation of an instance of the modelling code, prepared as described in Section \ref{sec:model_preparation}, which the system will then automatically schedule for execution against a particular semantic context. At model deployment, the user indicates a desired model name, the modelling code to be used (a package name, a version and a model class) and the target context (identified by signal and entity) to which the model should be applied. The user can also specify a training (scoring) deployment, indicating a time when the model should start training (scoring) and a training (schedule) frequency. Additional custom user parameters can also be included in the model deployment and will be made available to the modelling code upon execution, offering further customisation. Listing \ref{alg:model_deployment_example} shows an example of model deployment in JSON format. As discussed in Section \ref{sec:model_preparation}, the model deployment parameters will be made available to the corresponding model implementation, as in Listing \ref{alg:model_implementation_example}, at execution time. As explained in Section \ref{sec:overview}, the system takes care of automatically executing the deployed models at the defined training and scoring schedules, and of automatically persisting model version data (trained model parameters) and time-series prediction. The serverless cloud computing framework is leveraged for built-in parallelization and (theoretically) infinite horizontal scalability. 

Separating the model implementation from the deployment configuration creates a
flexible and powerful system for managing large numbers of models
programmatically. New deployments of an existing model can be generated for any number of target semantic contexts, without the need to customise or write new modelling code. This is a very powerful feature, particularly in large and complex IoT applications. It is possible, for example, to create a simple routine that explores the semantic representation of the application and automatically deploy models based on desired semantic rules, such that a forecasting application adapts and grows as new IoT sensors are added to the system.  

The system supports the deployment of any desired number of models for a given semantic context. It is often the case that there might not be one particular AI modelling technique or AI model architecture that is clearly superior for a specific time-series prediction task. The user can therefore decide to compare a set of models for the task. Ensemble models that receive as input the predictions from other models are also naturally supported by the modelling framework. A model ranking mechanism also exists whereby the user can specify the order of priority of the models. This can be used to serve the best prediction to downstream applications which retrieve data only based on the semantic context and do not need to know which specific model (or set of models) was used to generate the prediction.

\section{Smart-grid forecasting deployment} \label{sec:results}

IBM Research Castor was demonstrated in three smart-grid scenarios across Europe, specifically in Switzerland, Germany and Cyprus, as part of GOFLEX, a research project funded by the European Union \cite{goflex}. The deployments were aimed at the provisioning of large amounts of localised short-term time-series forecasts of distributed energy demand and renewable generation, to support decisions in market-based energy-flexibility trading. The following Section \ref{sec:results_data} details the scale and complexity of the IoT data handled by the system, and describes the use of the modelling framework to deploy necessary data transformations. The deployment of time-series AI models for energy prediction is then demonstrated in Section \ref{sec:results_modelling}. The scalability of the system is discussed in \ref{sec:results_scalability}.

\subsection{Data Ingestion and Transformation} \label{sec:results_data}

\begin{figure}
    \centering
    \includegraphics[width=8cm]{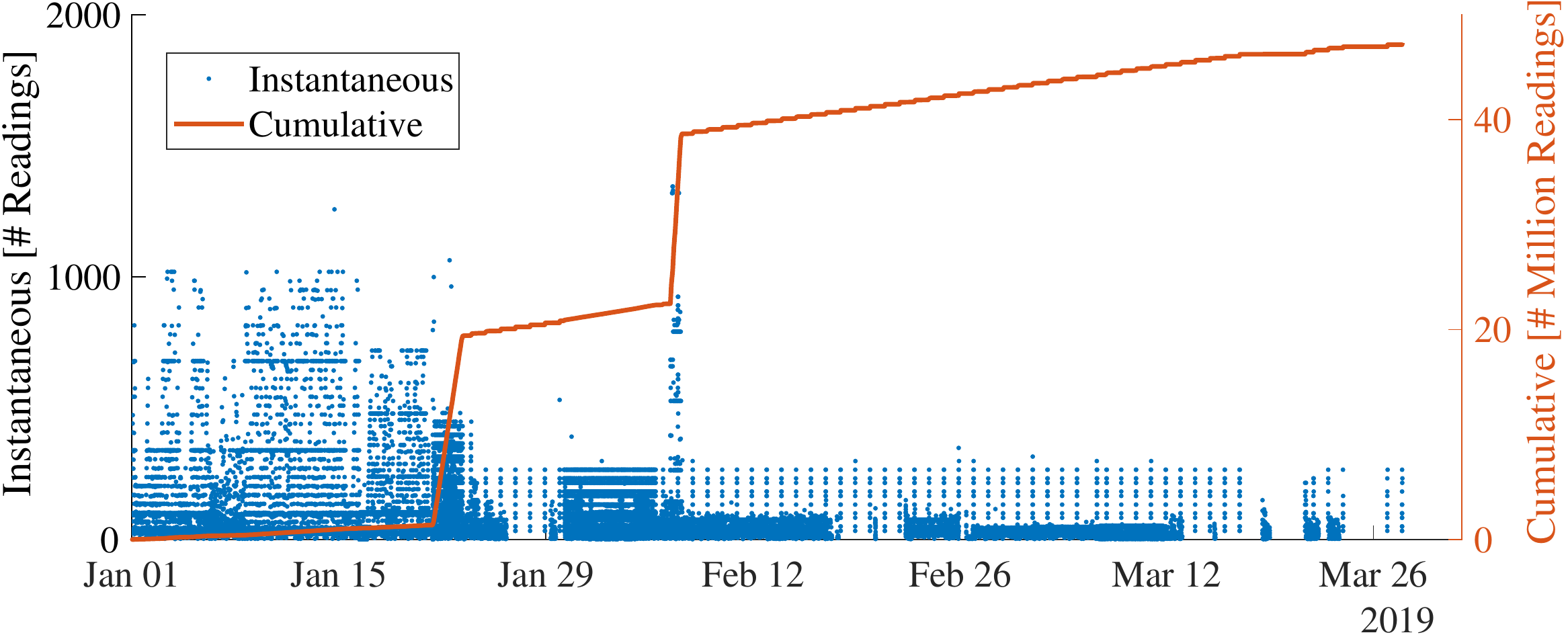}
    \caption{IoT data ingestion from the Cyprus smart-grid site of the GOFLEX project \protect\cite{goflex}, with 500 sensors sending nearly 15 million readings per month.}
    \label{fig:cyprus-ingestion}
\end{figure}

In each smart-grid scenario, data were received from several IoT sensors. Examples of data sources included smart meters at residential and industrial prosumers (consumers and producers of electricity), distribution system operator SCADA systems and energy billing systems.  Specific quantities were voltage, current, power consumption/generation, grid asset load and energy market data. Figure \ref{fig:cyprus-ingestion} shows the rate of data ingestion at the Cyprus site from January through March 2019, where on average, 15 million readings were received each month (nearly 1.4K per hour) from about 500 sensors.  

\begin{figure}
    \centering
    \includegraphics[width=8cm]{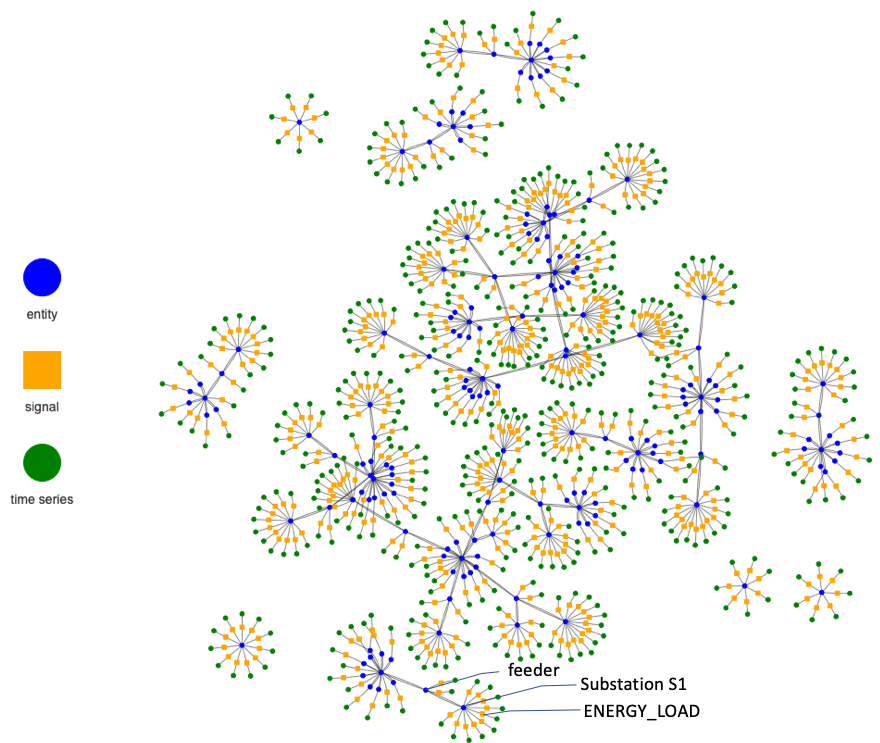}
    \caption{Semantic representation of the IoT data.}
    \label{fig:cyprus-context}
\end{figure}

As detailed in sections \ref{sec:overview} and \ref{sec:modelling}, the system manages the IoT data in a knowledge-based time-series store. Figure \ref{fig:cyprus-context} shows a portion of the semantic graph associated with the data ingested from the Cyprus site: every sensor is represented by a time-series node in the graph, which is then connected to nodes expressing the semantic concepts of signal (e.g. energy import/export, voltage magnitude) and entity (e.g. prosumers, substations, feeders); topology between entities is also represented (e.g. which feeder of which substation a given prosumer is connected to). As explained in Section \ref{sec:model_preparation}, the context of the data is available to the model code, and can be conveniently leveraged for generalising feature selection and engineering tasks based on abstract semantic concepts.

\begin{figure}
    \centering
    \includegraphics[width=8cm]{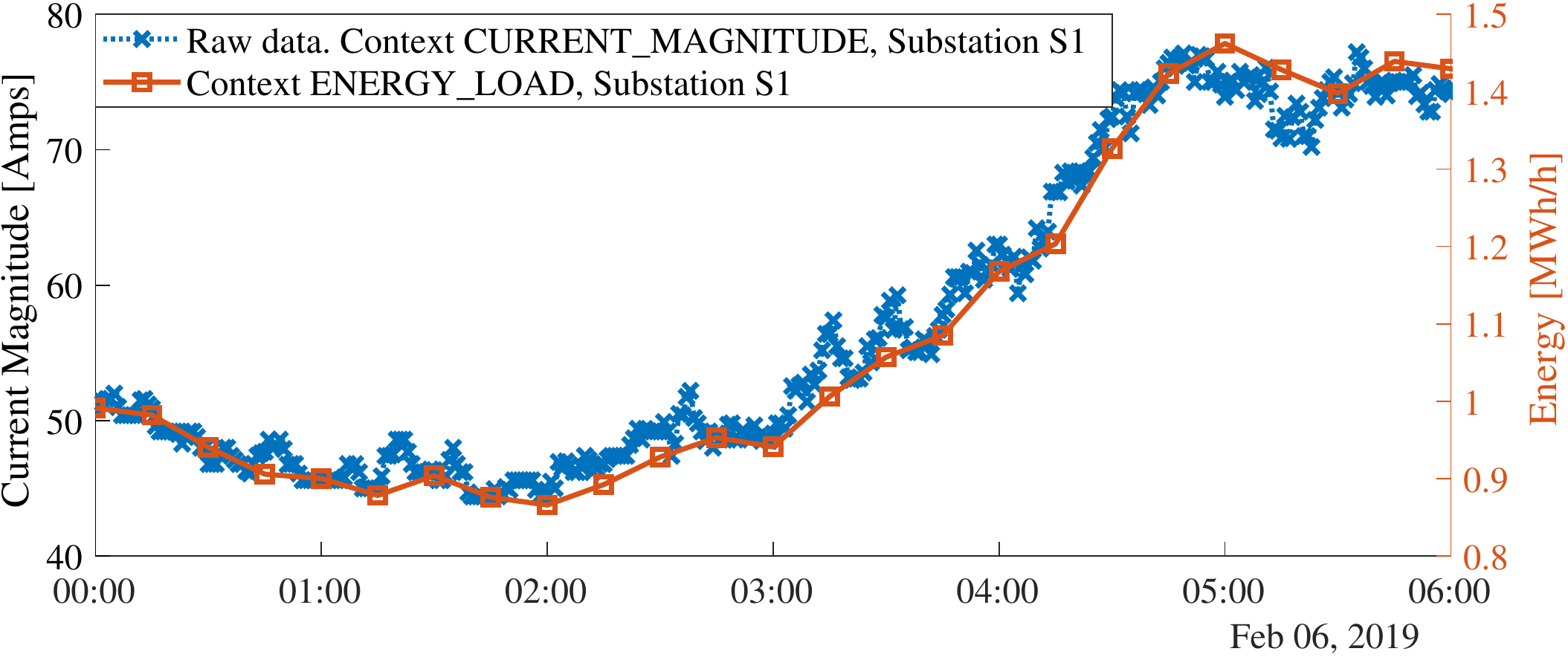}
    \caption{Example of data transformation model, integrating instantaneous current magnitude to produce estimate of energy load.}
    \label{fig:cyprus-sample-transform}
\end{figure}

Beside their spatial complexity, raw sensor-data are not always received at a regular or consistently aligned time resolution. Also, in some cases, time-series data for a quantity of interest for prediction are not observed directly but need to be computed from aggregation and transformation of other time-series. The modelling framework outlined in Section \ref{sec:modelling} can be conveniently used to express data transformation models. Figure \ref{fig:cyprus-sample-transform} shows an example where raw sensor data of instantaneous current magnitude at one-minute time resolution are scaled and integrated to obtain a time-series of energy sampled at 15-minute resolution, since that is the target for energy forecasting. To the user of the system, the time-series produced but the transformation model appears as any other raw time-series and can be retrieved using semantics.  

\subsection{AI Modelling} \label{sec:results_modelling}

Machine learning models have been developed based on the data received from the IoT sensors, outlined in Section \ref{sec:results_data}, with the specific objective of delivering energy predictions over a 24-hour horizon. 
In the following, the developmend and deployment of AI models to predict energy demand at a distribution substation is discussed. 
Linear regression (LR), general additive model (GAM), artificial neural network (ANN) and long short-term memory (LSTM) artificial neural network models were selected to represent a wide spectrum of the most common techniques used in energy demand forecasting. The features used for the four models are defined in Table~\ref{tab:model_features}. 
 
\begin{table}[ht]
    \centering
    \begin{tabular}{| p{1cm} | p{6.5cm}|}
    \hline
    \textbf{Model}     & \textbf{Features Overview}\\
        \hline
         LR & Weather forecasts (temperature), lag features (weather and target at 1- to 24-hour lags), calendar features (time-of-day, week-day) \\
        \hline
         GAM  & Weather forecasts (temperature), lag features (weather and target at 1- to 24-hour lags), calendar features (time-of-day, week-day)  \\
        \hline
         ANN & Weather forecasts (temperature), lag features (target at 1- to 192-hour lags) \\
        \hline
         LSTM & lag features (target at 1- to 24-hour lags) \\
        \hline
    \end{tabular}
    \caption{AI Model features selected.}
    \label{tab:model_features}
\end{table}

The training data consisted of hourly data of energy demand from January 1, 2018 to January 1, 2019 along with the corresponding weather data at that location. The models were validated on the subsequent hourly values until February 5, 2019. The ANN model architecture consists of 4 layers with 512 hidden neurons with Rectified linear unit (ReLu) activation and one output layer with sigmoid activation. The LSTM model has a similar architecture, but with 2 hidden layers. Both model weights were fitted using Adam Otimizer, with a learning rate of $0.001$. Batch sizes and epochs were chosen for each model based on cross-validation.         

The models were prepared by implementing the $load-transform-train-score$ functions, following the modelling framework outlined in Section \ref{sec:model_preparation}. 
The model code can make use of parameters that will be populated at execution (train or score) time from the system based on the specifics of a model deployment, as detailed in Section \ref{sec:model_deployment}. In particular, the semantic context parameter was used in the $load$ function (refer to Listing \ref{alg:model_implementation_example})  to retrieve the relevant weather feature data (based on entity GIS coordinates) and the historical target time-series data (based on the context signal, entity) required for training or for preparing lagged features. The $train$ function was implemented to learn the machine-learning model on historical data and return a model version object containing the fitted parameters. The $score$ function was implemented to compute and return model predictions over a 24-hour horizon. The additional  $user\_parameters$ deployment configuration parameter was used to control the training/scoring time windows and the time resolution of the predictions. 

\begin{figure}[ht]
    \centering
    \includegraphics[width=7cm]{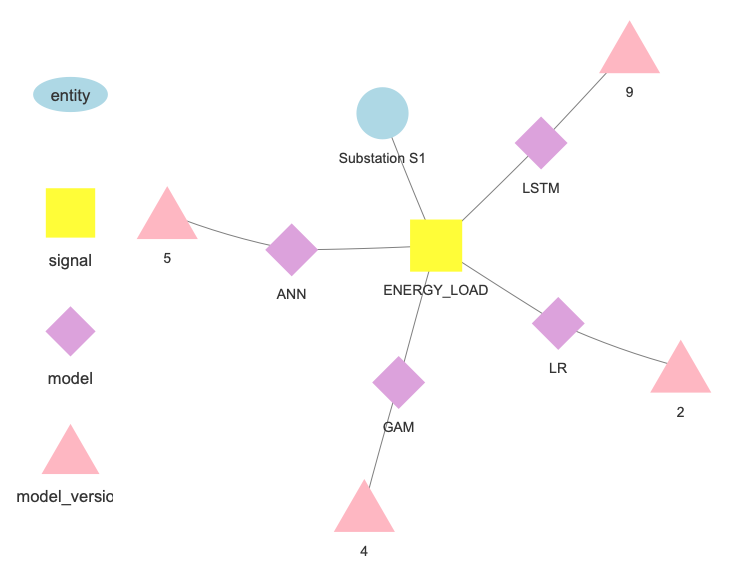}
    \caption{Deployed models and trained versions for a context.}
    \label{fig:model_hierarchy}
\end{figure}

\begin{figure}[ht]
    \centering
    \includegraphics[width=8cm]{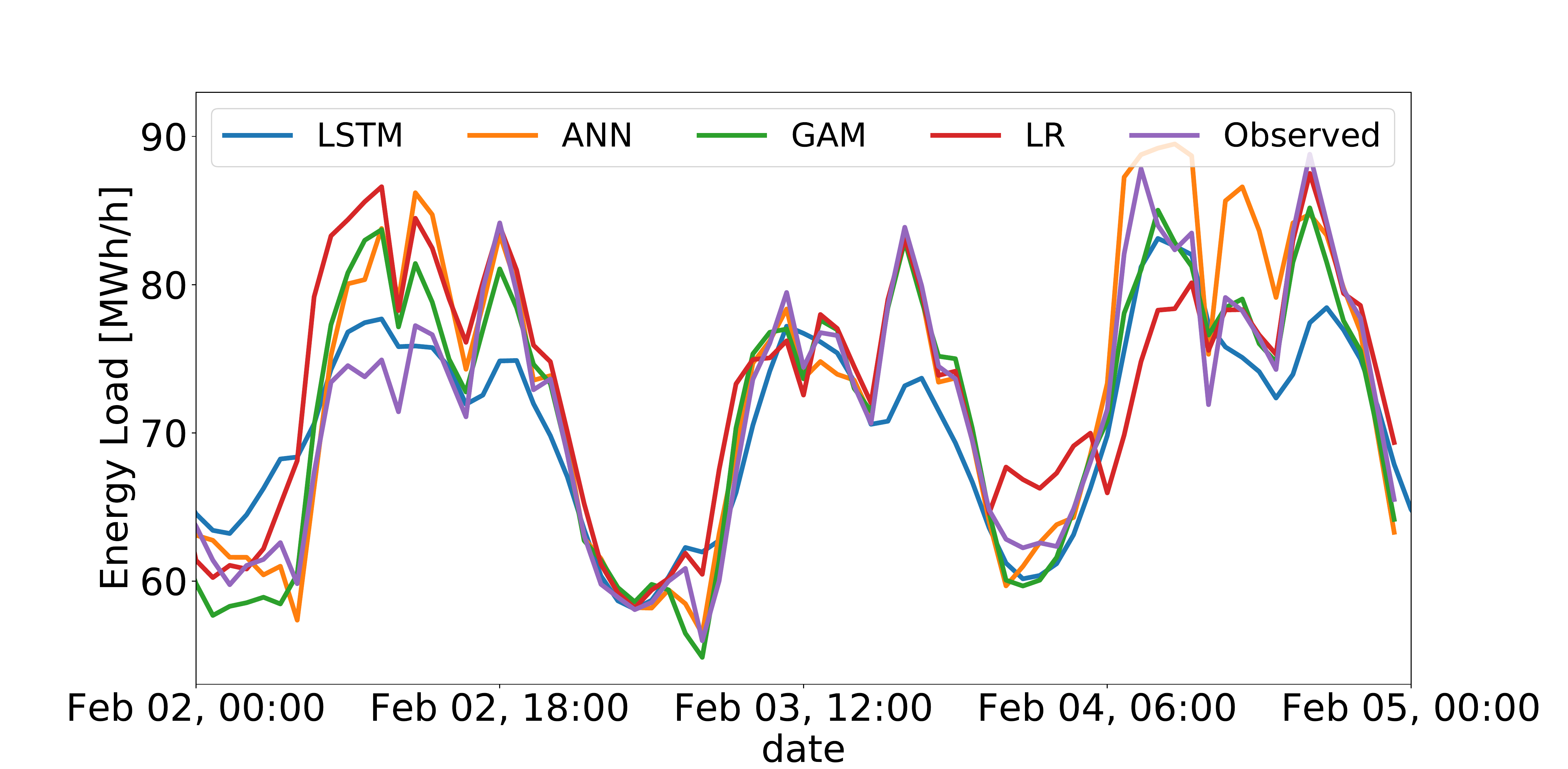}
    \caption{Forecasts from multiple models for a single context.}
    \label{fig:models}
\end{figure}

\begin{figure}[ht]
    \centering
    \includegraphics[width=8cm]{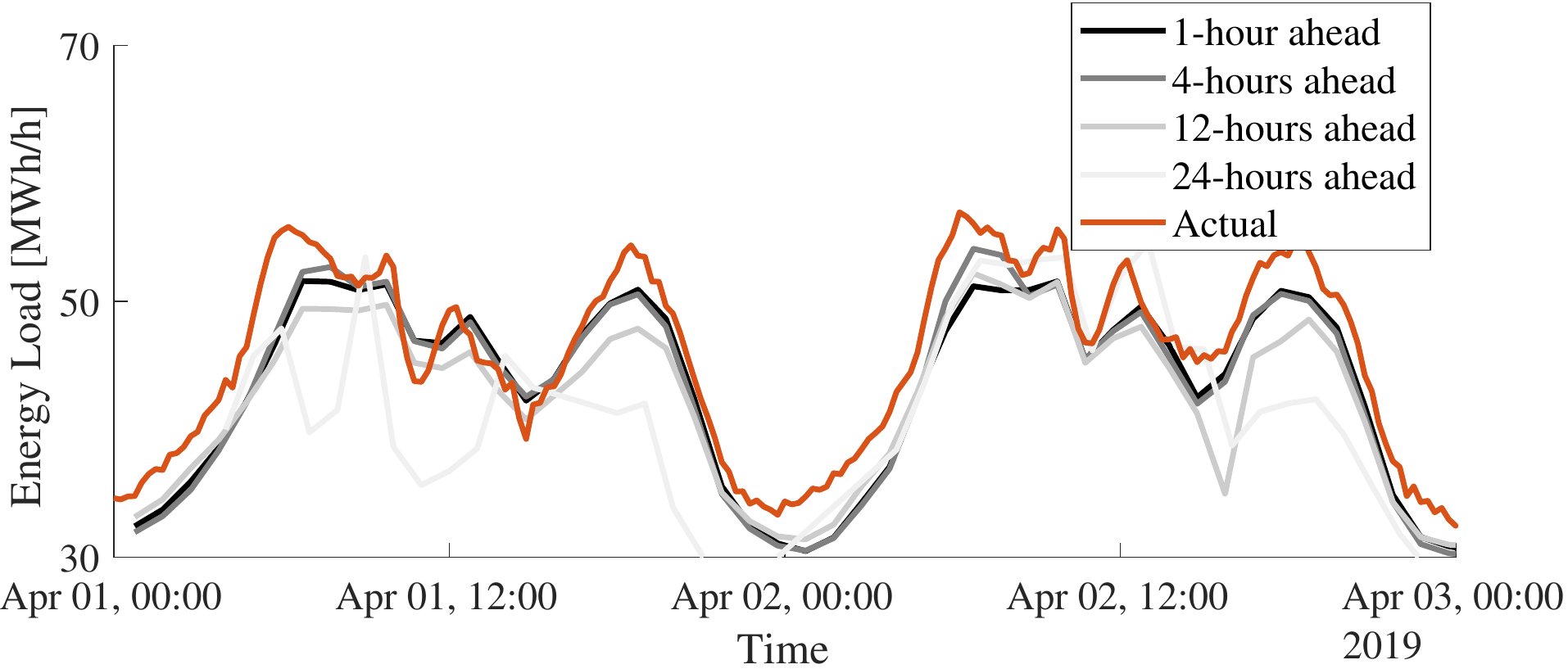}
    \caption{Time-series predictions, from the GAM model of Fig. \ref{fig:models}, with different forecasting horizons.}
    \label{fig:model_horizon}
\end{figure}

Specific model instances were deployed by preparing model deployment configurations, as outlined in Section \ref{sec:model_deployment}. The semantic context was set to the target entity (e.g. $SUBSTATION$ $S1$) and signal (e.g. $ENERGY\_LOAD$), the training and scoring schedules were defined (hourly and weekly, respectively), custom user parameters were used to specify a training window, the time and the desired forecasting window. After deployment, the system automatically executed the models, by training model versions and populating time-series predictions, as explained in Section \ref{sec:overview}. The models deployed for a given context can be explored in the semantic graph, as shown in Fig. \ref{fig:model_hierarchy}, and the time-series predictions can be retrieved and compared to the actual measurements (coming from raw or transformed sensor data), as shown in Fig. \ref{fig:models}, leveraging semantic-based APIs. Model performance is therefore easily tracked. 
In the example, the observed mean-absolute-percentage error (MAPE) for LR, GAM, ANN, LSTM on the validation period is $3.92\%$, $2.86\%$, $2.76\%$ and $6.37\%$ respectively. 
It is important to emphasize, here, that the accuracy of the models depends on multiple factors including models architecture, choice of features, hyperparamethers and even size of historical data. 
Identifying the optimal modelling technique and model setup for the considered prediction task goes beyond the scope of this paper. A grid search can for example be performed to optimize the proposed models. The interested reader is referred to the work of \cite{bouktif2018optimal} and references within for implementation details.   

Since the full history of predictions is persisted and rolling-window forecasts are not overwritten, the historical performance of a predictive model can also be validated across multiple prediction horizons, as shown in Figure \ref{fig:model_horizon}. 


\subsection{System Scalability} \label{sec:results_scalability}


Table \ref{tab:deployed_systems} summarises the current scale of the three smart-grid deployments in terms of quantity of sensors and deployed models (both AI models and data transformation models). Note that the figures in Table \ref{tab:deployed_systems} refer to the initial set up at the three sites (at the time of writing this paper), which are planned to be expanded by about one order of magnitude in the near future. The deployment of hundreds of models was semi-automated by programmatically setting the deployment configurations for the AI model implementations defined in Section \ref{sec:results_modelling} to all required semantic targets. For example, the 174 models of site 3 are based on only 6 model implementations. The average recorded model execution (scoring task), in Table \ref{tab:deployed_systems}, shows how new forecasts are available within less than 20s from being triggered. As detailed in Section \ref{sec:overview}, model execution leverages the serverless cloud computing framework. The resources allocated to the individual serverless jobs was 2 CPUs and 2GB of RAM memory. The resource specification of the knowledge-based time-series services were 1 CPU Core, 4GB RAM for the relational database and 1 CPU, 0.5GB RAM for the graph database. 

\begin{table}[ht]
    \centering
    \begin{tabular}{l|ccc}
    \hline 
    \rule{0pt}{2ex} Site &\# Sensors &\# Models & Execution [s]\\
    \hline 
     Germany &18  &11 &16.8\\
     Switzerland &196 &61 &19.7 \\
     Cyprus &531 &174 &15.9\\
    \hline
    \end{tabular}
    \caption{Size and performance in deployed systems. Model execution refers to the average duration of a scoring job.}
    \label{tab:deployed_systems}
\end{table}

Experiments were designed to assess system scalability with respect to the numbers of AI models. The analysis was focussed on the model scoring task, since that is the most time-critical and frequent job in an IoT system. An increasing number of model scoring jobs (using the GAM model discussed in section \ref{sec:results_modelling}), ranging from 10 to 200, were launched in parallel and each duration was recorded. The average duration and the projected number of models that can be executed in one hour are summarised in Table \ref{tab:scalability}. As the number of parallel executions is increased from 10 to 50, the average duration is only marginally affected, from 6.4s to 9.5s, so that the system is able to perform an almost theoretical 4-fold higher number of model scoring jobs in one hour, from 5.6K up to 20K. A diminishing gain is observed, however, with 150 parallel jobs producing about 27K jobs per hour. The current system setup is optimal at 175 jobs in parallel, with 200 parallel jobs giving no appreciable performance gain.

The observed flattening of the model performance is mainly due to practical resource limitations in the back-end database services used by the time-series micro-services, which are queried at model execution for loading model data (parameters, configuration), loading time-series data and saving the model predictions (refer to the system workflow in Fig. \ref{fig:system_overview}). Investing in additional resources on these back-end services, based on a trade-off between application requirements and cloud infrastructure costs, will provide further gains in the scalability of modelling jobs. 

\begin{table}[ht]
    \centering
    \begin{tabular}{rrr}
    \hline 
    Parallel Jobs & \# Jobs/hour & Job Duration [s]\\
     \hline 
    10 & 5,600 & 6.4 \\
    50 & 18,900 & 9.5 \\
    100 & 22,300 & 16.1 \\
    150 & 26,900 & 20.1 \\
    175 & 27,600 & 22.8 \\
    200 & 26,700 & 27.0 \\
    \hline 
     \end{tabular}
    \caption{System scalability analysis.}
    \label{tab:scalability}
\end{table}

\section{Conclusion} \label{sec:conclusion}

IBM Research Castor, a novel system for managing and deploying AI models on the cloud, was described. The system is particularly suitable for time-series forecasting in large-scale IoT applications, where both the size and the complexity of the data are a challenge. Results from a number of live deployments of the system in real-world smart-grid forecasting scenarios were discussed. It was shown how some of the typical machine-learning models can be easily prepared and deployed for automatic training/scoring on the cloud. Features for handling the ingestion from IoT sensor data, the knowledge-based contextualization of the time-series data and data transformation were also demonstrated. Scalability of the system in terms of running tens of thousands of models was also analysed.

\section*{Acknowledgments}
This research has received funding from the European
Research Council under the European Unions Horizon 2020
research and innovation programme (grant agreement no.
731232).

\bibliographystyle{named}
\bibliography{goFlexBib}

\end{document}